\documentclass[reprint,twocolumn,floatfix,superscriptaddress,showpacs]{revtex4-1}
\usepackage{xfrac}
\usepackage{amsfonts}
\usepackage{float}
\usepackage{dsfont}
\usepackage{amsmath}
\usepackage{amssymb}
\usepackage{amsthm}
\usepackage{mathtools}
\usepackage{graphicx}
\usepackage{color}
\usepackage{changes}
\usepackage{bm}
\usepackage{bbm}
\usepackage[normalem]{ulem}
\usepackage[hidelinks]{hyperref}
\hypersetup{
     colorlinks   = true,
     citecolor    = blue,
     linkcolor    = blue
}
\usepackage{mathptmx}
\usepackage{times}
\usepackage{enumerate}

\newcommand{\ignore}[1]{}
\newcommand{\nobibentry}[1]{{\let\nocite\ignore\bibentry{#1}}}

\newcommand{\ket}[1]{\left\vert#1\right\rangle}

\def\ketbra#1#2{{\vert#1\rangle\!\langle#2\vert}}

\definecolor{amber}{rgb}{1.0, 0.75, 0.0}

\DeclareMathAlphabet{\mathcal}{OMS}{cmsy}{m}{n}

\theoremstyle{definition}

\theoremstyle{plain}

\theoremstyle{plain}

\usepackage{graphicx}
\usepackage{braket}
\usepackage{dsfont}
\usepackage[english]{babel}
\usepackage{amssymb, amsmath}
\usepackage{amsthm}
\usepackage{xcolor}

\newcommand{\be}{\begin{equation}} 
\newcommand{\bes}{\begin{equation*}} 			% math environments
\newcommand{\ee}{\end{equation}}
\newcommand{\ees}{\end{equation*}}

\newcommand{\bematrix}{\left(\begin{matrix}}
\newcommand{\ematrix}{\end{matrix}\right)}
\begin{document}

\title{Robustness of non-locality in many-body open quantum  systems}
\date{24/02/2022}
\author{Carlo Marconi}
\affiliation{F\'isica Te\`{o}rica: Informaci\'o i Fen\`{o}mens Qu\`{a}ntics.  Departament de F\'isica, Universitat Aut\`{o}noma de Barcelona, 08193 Bellaterra, Spain}
\author{Andreu Riera-Campeny}
\affiliation{F\'isica Te\`{o}rica: Informaci\'o i Fen\`{o}mens Qu\`{a}ntics.  Departament de F\'isica, Universitat Aut\`{o}noma de Barcelona, 08193 Bellaterra, Spain}
\affiliation{Institute for Quantum Optics and Quantum Information of the Austrian Academy of Sciences, A-6020 Innsbruck, Austria}
\affiliation{Institute for Theoretical Physics, University of Innsbruck, A-6020 Innsbruck, Austria}
\author{Anna Sanpera}
\affiliation{F\'isica Te\`{o}rica: Informaci\'o i Fen\`{o}mens Qu\`{a}ntics.  Departament de F\'isica, Universitat Aut\`{o}noma de Barcelona, 08193 Bellaterra, Spain}
\affiliation{ICREA, Pg. Lluis Companys 23, E-08010 Barcelona, Spain}
\author{Albert Aloy}
\affiliation{Institute for Quantum Optics and Quantum Information - IQOQI Vienna, Austrian Academy of Sciences, Boltzmanngasse 3, 1090 Vienna, Austria}

\begin{abstract}
\noindent Non-locality consists in the existence of non-classical correlations between local measurements. So far, it has been investigated mostly in isolated quantum systems. Here we show that non-local correlations are present, can be detected and might be robust against noise also in many-body open quantum systems, both in the steady-state and in the transient regime.
We further discuss the robustness of non-local correlations when the open quantum system undergoes repeated measurements, a relevant scenario for quantum cryptography.
\end{abstract}

\maketitle

\textit{Introduction}.---Non-locality refers to the existence of correlations between local measurements on a shared resource, that cannot be explained by any local realistic theory \cite{bell1964einstein}. Besides their importance from a foundational point of view, such correlations are becoming increasingly relevant as a resource for secure quantum technologies involving device-independent (DI) quantum key distribution \cite{ekert1991quantum,barrett2005no,scarani2006secrecy,acin2007device}, shared randomness certification and amplification \cite{pironio2010random,colbeck2012free} or quantum communication techniques \cite{relativisticbell}. Operationally, non-locality is assessed by means of Bell inequalities (see e.g. \cite{brunner2014bell}).

It is well known that entanglement and non-locality
are related but inequivalent resources \cite{Brunner2005, vertesi2014disproving,augusiak2015entanglement,friis2019entanglement}: while every non-local state is entangled, the converse is not necessarily true. Within the context of many-body systems, the analysis of entanglement has become a major trend of modern physics leading to seminal insights in many areas. On the contrary, the role of non-local correlations in such systems remains widely unexplored, with few significant advances \cite{tura2014detecting,tura2017energy}.
The reasons behind this gap are several. First, the characterization of non-locality in many-body systems typically requires the construction of $N$-body correlators \cite{werner2001all,zukowski2002bell}, an approach which is, in general, unfeasible within the current technological capabilities. 
Second, the complexity of this task grows with the dimension of the system, resulting, generically, in an NP-complete problem \cite{babai1991non}.
Nevertheless, recent progress concerns the construction of many-body Bell inequalities that are constrained by symmetries and involve only one- and two-body correlators \cite{tura2014detecting,tura2015nonlocality,aloy2019device,piga2019bell}.

All such examples refer to isolated systems, and it is still an open question whether the interaction with an environment would result in a decay of non-local correlations. Besides of its theoretical implications, answering to this question would bring key insights at an operational level, especially in view of the recent DI proposal to certify quantum information tasks without requiring trust on their implementation.

Even though one is tempted to believe that non-locality, due to its fragile nature, is lost when a system interacts with an environment, here we prove, by detecting violation of Bell inequalities in many-body open quantum systems (OQS), that this is not necessarily the case.

To this aim, we choose the Lipkin-Meshkov-Glick (LMG) model \cite{LMG1} as our system of interest, \text{S}, and address, by means of a quantum master equation (QME), its dynamical evolution when it is put in contact with an external environment,  \text{E}. The latter typically acts as an infinite bath/reservoir at inverse temperature $\beta = 1 / T (\kappa_B = 1)$, but we investigate as well the effect of other dissipators.  QMEs \cite{breuer2002theory} provide an efficient tool to describe the evolution of \text{S}, as well as its non-equilibrium steady-states (NESS). To certify the presence of non-locality, we make use of Bell inequalities specifically designed for symmetric many-body systems \cite{tura2015nonlocality}. Finally, we address an adversarial scenario in which a non-local many-body system undergoes repeated measurements and we discuss the robustness and survival time of non-local correlations in this setup.

\textit{Bell inequalities}.---The simplest multipartite Bell experiment consists of $N$ spatially separated observers, each of them able to perform two dichotomic measurements. The experiment proceeds as follows: i) an $N$-partite resource (a  multipartite quantum state) is shared among $N$ parties; ii) each party, $\mathcal{O}_i$, performs a measurement $\mathcal{M}_r^{(i)}$ on the shared resource, where $r\in\{0,1\}$ labels the two measurement settings; iii) each measurement yields an outcome $a_i\in\{0,1\}$. This procedure is repeated several times until a sufficient amount of statistics has been gathered. Then, the statistics collected through the experiment can be described in terms of the statistical correlations that depend on the number of parties, their input measurement settings and the obtained outcomes. A standard Bell inequality is a linear combination of such correlations, and its bound denotes the limit given by all the correlations that can be observed under the physical principles of locality and realism. Hence, the violation of a Bell inequality certifies the presence of non-locality.\\
A caveat of standard multipartite Bell inequalities is that they typically rely on expectation values that involve many parties, while requiring the individual addressing of each party. In this work, we restrict our analysis to a subclass of Bell inequalities, dubbed as \textit{two-body permutationally invariant} \cite{tura2014detecting,tura2015nonlocality}, which are constrained by symmetries and involve at most two-body correlation functions, i.e., \cite{tura2015nonlocality}
\begin{equation}
\label{bell}
\mathcal{B}(\phi_i,\theta_i) = 2N \mathds{1}_{2^N} - 2 \mathcal{C}_0  + \frac{1}{2}   \mathcal{C}_{00}  -  \mathcal{C}_{01}  + \frac{1}{2}   \mathcal{C}_{11}~,
\end{equation}
where $\mathcal{C}_{r} = \sum_{i=0}^{N-1}  \mathcal{M}^{(i)}_{r},~
 \mathcal{C}_{rs}  = \sum_{{i \neq j=0}}^{N-1}   \mathcal{M}^{(i)}_{r}\mathcal{M}^{(j)}_{s}$, are the one- and two-body symmetric correlators respectively, with measurement settings $r,s \in \{0,1\}$. We consider the case when all the parties perform the same measurements, i.e., $\theta_{i}=\theta_{j}=\theta$, $\phi_{i}=\phi_{j}=\phi$ for all $i \neq j$, since it is conjectured in the literature that the maximal violation associated to the Bell operator of Eq.(\ref{bell}) is achieved by this choice of the measurement settings \cite{tura2015nonlocality,aloy2019device,piga2019bell}.
Here $\mathcal{M}^{(i)}_{0} = \cos(\phi) \sigma^{(i)}_{z} + \sin(\phi) \sigma^{(i)}_{x}$ and $
\mathcal{M}^{(i)}_{1} = \cos(\theta) \sigma^{(i)}_{z} + \sin(\theta) \sigma^{(i)}_{x}$, where $\sigma^{(i)}_{\mu}$ denotes the Pauli matrix at site $i$ along the direction $\mu \in \{x,y,z\}$ and $(\phi,\theta)$ are the measurement angles of each party. Non-locality is revealed whenever $\mbox{Tr}[\mathcal{B}(\phi, \theta)\rho] <0$, certifying that the state $\rho$ is non-local and thus entangled. 

Eq.(\ref{bell}) is of particular interest when probing non-locality in symmetric $N$-qubit states, $\rho_{SYM}$, since the two-body reduced density matrix, $\rho_2=\mbox{Tr}_{1,\dots,N-2}(\rho_{SYM})$, is the same regardless of which $N-2$ systems have been traced out. If $\rho_{SYM}$ is represented in the symmetric subspace of $N$-qubits by means of the Dicke states, i.e., $\ket{D^{N}_{k}} = \binom{N}{k}^{-1/2} \sum_{\pi} \pi(\ket{0}^{N-k} \ket{1}^{k})$, where the sum runs over all the possible permutations $\pi$ of $N$ elements and $\{\ket{0},\ket{1}\}$ are the eigenstates of the Pauli matrix $\sigma_{z}$, then the two-body reduced density matrix reads
\cite{tura2015nonlocality}
\begin{equation}
\label{red}
(\rho_2)_{\mathbf{i}',~\mathbf{j}'} = \sum_{k=0}^{N-2} \frac{\binom{N-2}{k}\left( \rho_{SYM} \right)_{k + |\mathbf{i}'|,~k + |\mathbf{j}'|}}{\sqrt{\binom{N}{k + |\mathbf{i}'|}\binom{N}{k + |\mathbf{j}'|}}}~,
\end{equation} 
\noindent where $0 \le \mathbf{i}',~\mathbf{j}' < 2^{2} $, $\mathbf{i}' = i_{0} i_{1},~\mathbf{j}' = j_{0} j_{1} $ are the binary representations of the labels associated to the matrix entries and $|\mathbf{i}|', |\mathbf{j}|'$ are their Hamming weights, i.e., the number of ones in this representation.\\
\noindent Hence, the two-qubit Bell operator can be derived \cite{tura2015nonlocality}
\begin{align}
\label{bell2}
	\mathcal{B}_2 (\phi, \theta) &= 2N \left(\mathds{1}_{2} \otimes \mathds{1}_{2} \right)
	+ \frac{N}{2} \left[ -2 \left(\mathcal{M}_{0} \otimes \mathds{1}_{2} + \mathds{1}_{2} \otimes \mathcal{M}_{0}\right) \right]\\ \nonumber
	&+\frac{N (N-1)}{2} [ \mathcal{M}_{0} \otimes \mathcal{M}_{0} + \mathcal{M}_{1} \otimes \mathcal{M}_{1}\\ \nonumber
	&-(\mathcal{M}_{0} \otimes \mathcal{M}_{1} + \mathcal{M}_{1} \otimes \mathcal{M}_{0}) ]~,
	\end{align}
which greatly reduces the computational cost required to probe non-locality. Nonetheless, let us remark that in order to perform a fully device-independent Bell experiment, one should still address all the parties individually. 

\textit{System model}.---The choice of the model for the system \text{S} is motivated by the aforementioned desirable properties that the Bell inequality (\ref{bell2}) offers. We consider a particular case of the LMG Hamiltonian, i.e.,
\begin{equation}
\label{lmg2}
H_{S} = \frac{J}{N} S_{z}^2- h S_{x}~,
\end{equation}
\noindent where $J$ is the interaction energy scale, $h$ is a magnetic field applied along the $x$ direction and the collective spin operators are $S_\mu = \frac{1}{2} \sum_{i=1}^N \sigma^{(i)}_{\mu}$, with $\mu \in \{x,y,z\}$~.
Since $H_{S}$ depends only on collective spin properties, it admits a block decomposition of the form $H_{S} = \bigoplus_{M} H_{S}(M)$, where the sum runs over the total spin number $M = M_{min}, \dots, N/2-1, N/2$~, with $M_{min} \in \{0,1/2\}$, depending on whether $N$ is even or odd. 
The ground state of Eq.(\ref{lmg2}) corresponds to a Gaussian superposition of Dicke states, which is known to maximally violate the Bell inequality associated to the Bell operator of Eq.(\ref{bell2}) \cite{tura2014detecting}. 
Given the symmetrical nature of our Bell inequality, we restrict our analysis to the block of maximum total spin $ M=N/2$, where the global spin operators can be expressed in the Dicke basis \cite{tura2015nonlocality}.

\textit{Open quantum system dynamics}.---OQS can be treated by means of QMEs to investigate their steady-state solutions as well as their transient regime. Assuming that at $t=0$ the initial state of the total system can be written as $\rho(0) = \rho_{S}(0) \otimes \rho_{E}(0)$, the evolution of \text{S} is obtained by tracing out the environment \text{E}, i.e.,
\begin{equation}
\rho_\text{S}(t) = \text{tr}_\text{E}[\rho(t)]\coloneqq \mathcal{E}(t)[\rho_\text{S}(0)]~,
\end{equation}
 where $\rho(t) = U(t)\rho(0)U(t)^\dagger$ is the total state at time $t$, $U(t) = \exp(-i H t)$, and the full Hamiltonian reads $H = H_\text{S}+H_\text{E}+\lambda V$, where  $V$ denotes the interaction between \text{S} and \text{E}, and $\lambda$ is a coupling parameter.  $\mathcal{E}(t)$ is a completely-positive and trace-preserving map usually referred to as the evolution map. QMEs provide a linear equation of motion for $\rho_\text{S}(t)$ of the form 
 $\mathcal{E}(t) = \exp(\mathcal{L} t)$, where $\mathcal{L}$ is the so-called Lindbladian operator, so that the evolution of $\rho_\text{S}$ can be cast in the form of the well-known Gorini-Kossakowski-Sudarshan-Lindblad (GKSL) master equation  \cite{GoKoSu1976,Li1976}
\begin{equation}
\label{gksl}
\mathcal{L}[\rho_\text{S}] = -i[H_\text{S},\rho_\text{S}] + \sum_k \gamma_k \left( \mathcal{J}_k \rho_\text{S} \mathcal{J}_k^\dagger -\frac{1}{2}\{\mathcal{J}_k^\dagger \mathcal{J}_k,\rho_\text{S} \}\right),
\end{equation}
where $H_\text{S}$ is the system Hamiltonian, $\gamma_k\geq 0$ are the dissipation rates, and $\mathcal{J}_k$ are the jump operators.

Interestingly, a system that is weakly and continuously interrogated also obeys an equation of the form of Eq.(\ref{gksl}). Given a party Eve that repeatedly performs the measurement $\mathcal{M}$ with outcomes $m_k$ and eigenprojectors $\Pi_k$, the explicit form of the associated Lindbladian is
\begin{equation}
\label{measurement}
\partial_t \rho_\text{S} = -i[H_\text{S},\rho_\text{S}] + \kappa \left(\sum_k \Pi_k \rho_\text{S} \Pi_k^\dagger - \rho_\text{S} \right)
\end{equation}
where $\kappa$ is the measurement rate \cite{CrBaJePe2006}.

\textit{Results}.---Our aim is to prove, detect and study the robustness of non-locality in a many-body multipartite OQS.
%where S is a collection of $N$ spin-1/2. T
To this aim we distinguish two scenarios: first, we explore how dissipation affects the detection of non-locality; and second, we examine the robustness of non-local correlations in an adversarial scenario in which the many-body system, S, undergoes repeated measurements.\\
We start by considering the stationary regime of the LMG model in contact with a bosonic bath at inverse temperature $\beta$. Let us denote by $\{b_{k}\}$ the bosonic operators ( $[b_k,b^{\dagger}_{k'}]=\delta_{k,k'}$) and assume an interaction
of the form $V= S_{y} \otimes (b_{k} + b_{k}^{\dagger})$. 
%Using standard techniques \cite{breuer2002theory},
We derive the corresponding QME (see Eq.(\ref{gksl})), and find the stationary solutions by imposing $\mathcal{L}[\rho_{S}]=0$.\\
In the low-temperature limit, we find that the steady states correspond to thermal states given by a Gaussian superposition of Dicke states of the form \cite{tura2014detecting}
\begin{equation}
\label{gauss}
\rho_{S} \approx \ketbra{\psi_N}{\psi_N}~, \quad \ket{\psi_{N}}=\sum_{k=0}^{N} \psi^{N}_{k} \ket{D^{N}_{k}}~,
\end{equation}
with  amplitudes $\psi^{N}_{k} \approx (1/\sqrt[4]{2\pi \sigma^2})e^{-(k-N/2)^2 /4\sigma^2}$, for some variance $\sigma^2$.

\begin{figure}[ht]
	\includegraphics[width=8.6cm,height=13cm]{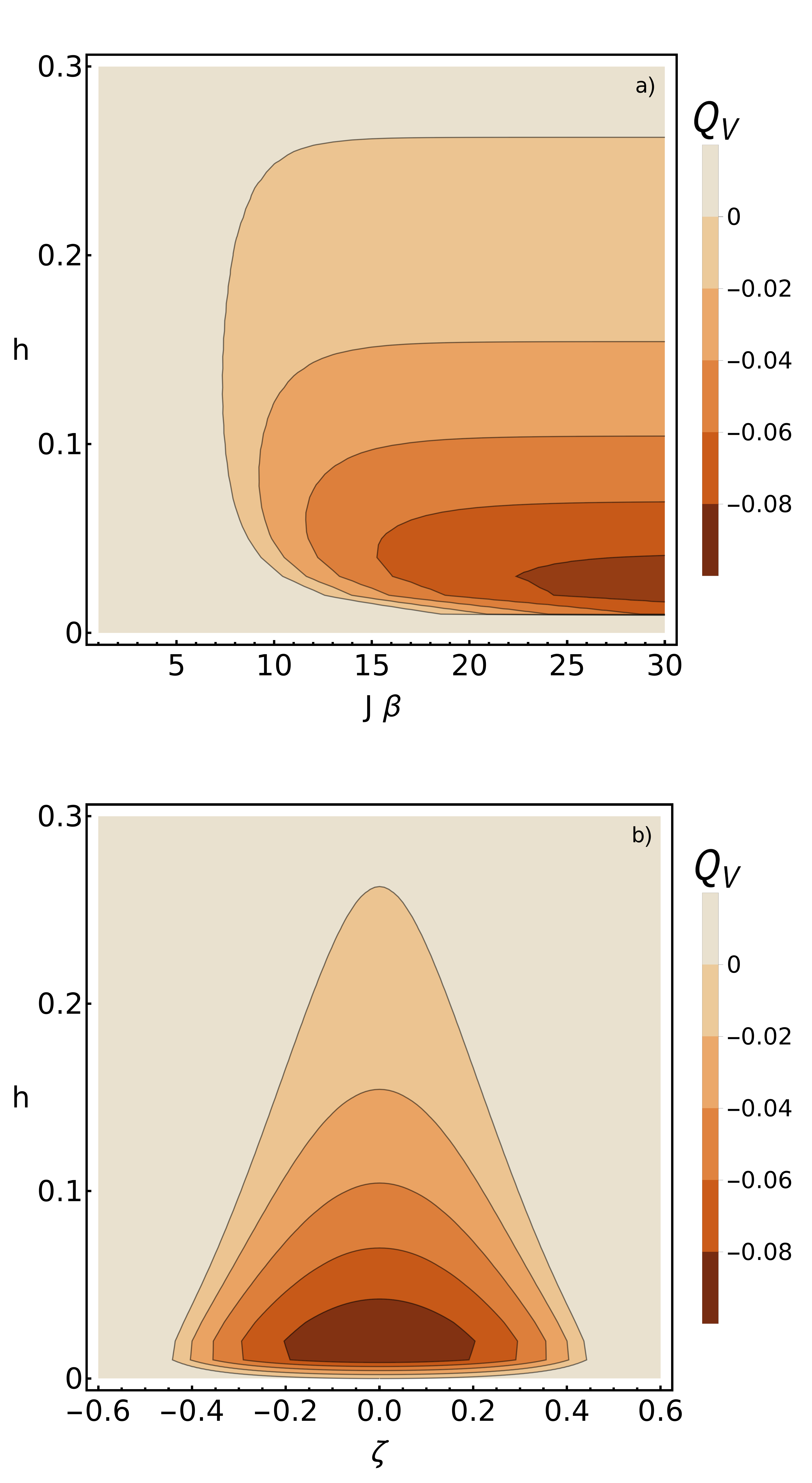}
	\caption{Non-local correlations in an OQS described by the Hamiltonian of Eq.(\ref{lmg2}), with $N=20$, in the case of: a) thermal noise, and b) a non-thermal jump operator $\mathcal{J}(\zeta)= \cos(\zeta) \hat{S}_{+} + \sin(\zeta) \hat{S}_{-}$. Negative values indicate the quantum violation ($Q_v$) of the Bell inequality associated to Eq.(\ref{bell2}).}
	\label{fig:ss}
\end{figure}
\noindent In Fig.\ref{fig:ss}a we display the detection of non-local correlations as a function of the magnetic field $h$, the energy coupling $J$ and the inverse temperature $\beta$. We observe that there is a significant region of the phase-space  for which $\mbox{Tr}[\mathcal{B}_{2}(\phi,\theta) \rho_{2}]<0$, providing evidence of their robustness against thermal noise. Notice that our results are consistent with the fact that, for $h=0$, the steady-states of $H_S$ of Eq.(\ref{lmg2}) are clearly separable. Moreover, for high temperatures, i.e., low values of $\beta$, thermal fluctuations hinder the possibility to observe non-locality. \\

Although the coupling with a thermal bath is the most common scenario when dealing with an OQS, we consider the effect of a dissipation that leads to non-thermal steady-states. To this end, referring to Eq.(\ref{gksl}), we design an \textit{ad hoc} jump operator of the form $\mathcal{J}(\zeta)= \cos(\zeta) \hat{S}_{+} + \sin(\zeta) \hat{S}_{-}$, where $\hat{S}_{\pm}=U^{\dagger} S_{\pm} U$ and $U$ is a unitary transformation from the Dicke basis to the energy basis, i.e., the set of eigenstates of $H_{S}$, and $S_{\pm}=S_x \pm i S_y$. Again, the resulting steady-state solutions are found by imposing $\mathcal{L}[\rho_{S}]=0$ for the corresponding GKLS master equation. In Fig.\ref{fig:ss}b we plot the non-local correlations detected by the Bell operator of Eq.(\ref{bell2}), and relate them to $J,h$ and the angle $\zeta$. Also in this case, we find that non-locality is present and can be detected for a large range of values of the parameter $\zeta$, showing to be robust against the effect of the magnetic field especially around $\zeta=0$. Indeed, notice that, for $\zeta = 0$, one gets $\mathcal{J}(0)=\hat{S}_{+}$, and the effect of the dissipation is to force the evolution towards a thermal steady state given by the same Gaussian superposition of Dicke states of Eq.(\ref{gauss}). However, for $\zeta \neq 0$, the steady-state solutions are not thermal states, a feature that makes our analysis in this regime particularly valuable.\\

While in Fig.\ref{fig:ss}a-Fig.\ref{fig:ss}b we showed that stationary steady-states exhibit non-local correlations, we are now interested in the somehow opposite regime of the dynamical evolution.
\begin{figure}[ht]
	\centering
	\includegraphics[width=8.6cm,height=10.05cm]{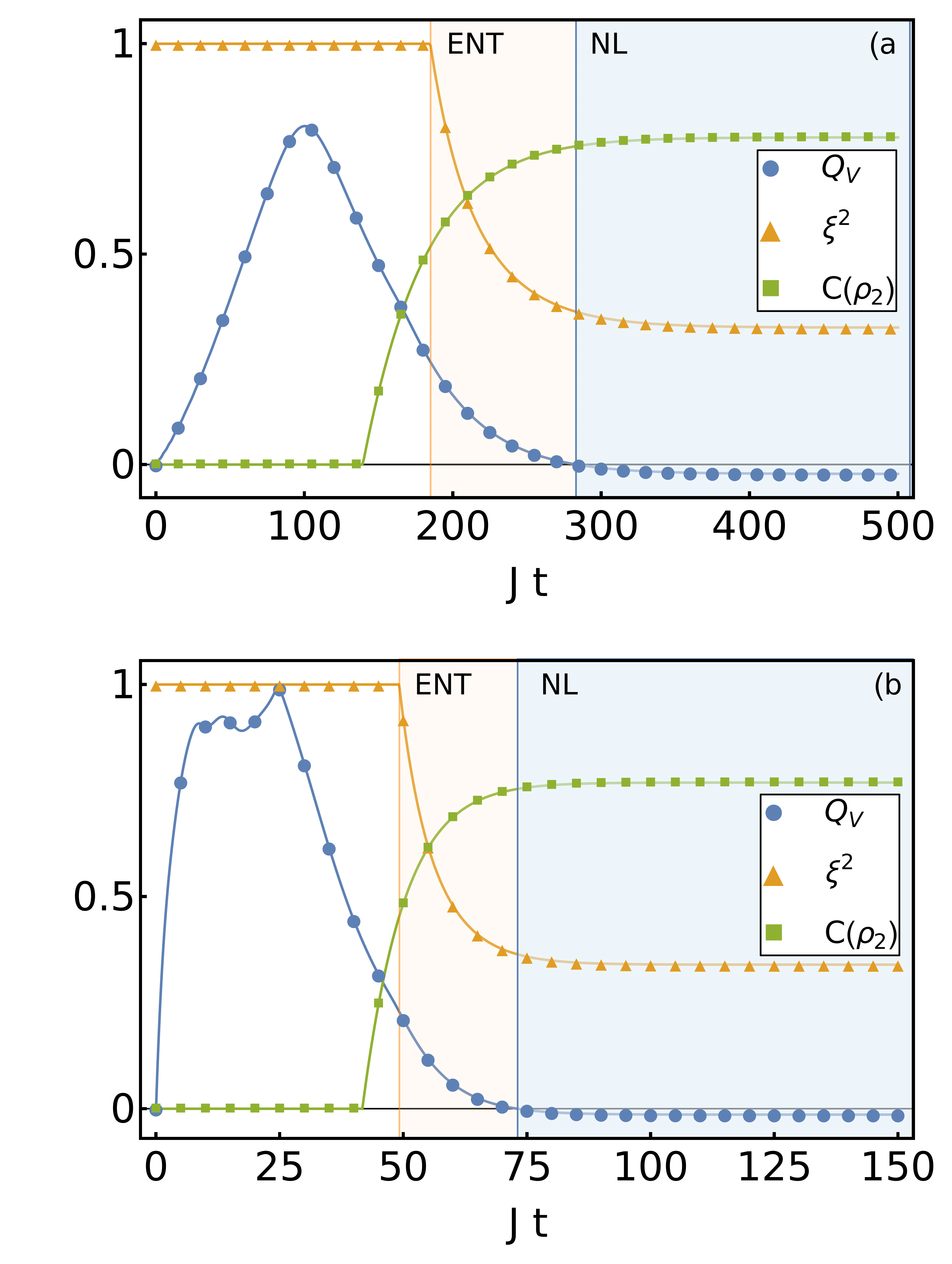}
	\caption{Comparison between concurrence (green squares), spin squeezing criterion (orange triangles) and non-local correlations (blue dots) detected by Bell operator (\ref{bell2}) in the state $\rho_{S}(t)$ for: a) thermal noise, and b) a non-thermal jump operator $\mathcal{J}(\zeta)= \cos(\zeta) \hat{S}_{+} + \sin(\zeta) \hat{S}_{-}$. The value of the violation in the stationary regime corresponds to the steady-state solutions of Fig.\ref{fig:ss} with $N=20,\; J=1,\; h=0.05$ for (a) $\beta =10$ and (b) $\zeta=0.35$.  
	NL (light blue) and ENT (orange) signal the region of non-locality and entanglement, respectively. Notice that the latter region comprises also the former. For the ease of readability, $C(\rho_{2})$ has been multiplied by a factor $N$ and $\xi^2$ has been chosen equal to 1 for separable states.}
	\label{fig:dy}
\end{figure}
In Fig.\ref{fig:dy}, we show how the stationary steady-state solutions of Fig.\ref{fig:ss} can be recovered as the result of a dynamical evolution. In particular, starting from a local initial state of the form $\rho_{S}(0)=\ketbra{D^{N}_{N}}{D^{N}_{N}}$, we show that non-local correlations can arise in the state $\rho_{S}(t)$, proving to be robust against the effect of dissipation both in the case of thermal and non-thermal noise. Furthermore, we compare the estimation of the entanglement of $\rho_{S}(t)$ provided by two distinct criteria. First, with the aid of Eq.(\ref{red}), we derive the two-qubit reduced density matrix $\rho_{2}$ and we compute its concurrence, $C(\rho_{2})$, defined as
\begin{equation}
C(\rho_{2})=\max(0,\lambda_{1}-\lambda_{2}-\lambda_{3}-\lambda_{4})~,
\end{equation}
\noindent where the $\lambda_{i}$'s are the eigenvalues of $\rho_{2}\Tilde{\rho}_{2}$, with $\Tilde{\rho}_{2} = \left(\sigma_{y}\otimes \sigma_{y}\right) \rho^{*}_{2} \left(\sigma_{y}\otimes \sigma_{y}\right)$, ordered in descending order.\\
Second, we inspect the value of the so-called spin squeezing parameter $\xi^{2}$, defined as
\begin{equation}
\xi^{2} = N \frac{(\Delta S_{z})^2}{\langle S_{x} \rangle^{2} + \langle S_{y} \rangle^{2}}~,
\end{equation}
\noindent where $(\Delta S_{z})^2 \equiv \langle S_{z}^{2} \rangle -\langle S_{z} \rangle^{2}$. In this case, if $\xi^{2}<1$, then the state is entangled. Notice that when dealing with symmetric states of $N$ qubits, the spin squeezing criterion is able to detect genuine multipartite entanglement \cite{wang2003spin}. 
In Fig.\ref{fig:dy} we display, for both the dissipators previously considered (i.e., thermal and non-thermal), the violation of the Bell inequality (\ref{bell2}) along with an estimation of the entanglement of $\rho_{S}(t)$ by means of the concurrence and the spin squeezing criterion. Notice that, in the case of thermal noise (Fig.\ref{fig:dy}a), non-local correlations manifest at a later time as compared to the case of the non-thermal dissipator (Fig.\ref{fig:dy}b). Interestingly, we observe a time gap also between the two estimations of the entanglement, a feature which is consistent with the different nature (i.e., two-body versus many-body) of the concurrence and the spin squeezing criterion. In both cases our results provide further evidence that non-locality and entanglement are inequivalent resources, as they clearly manifest at different times scales. \\

\begin{figure}[ht]
	\centering
	\includegraphics[width=8.6cm,height=7cm]{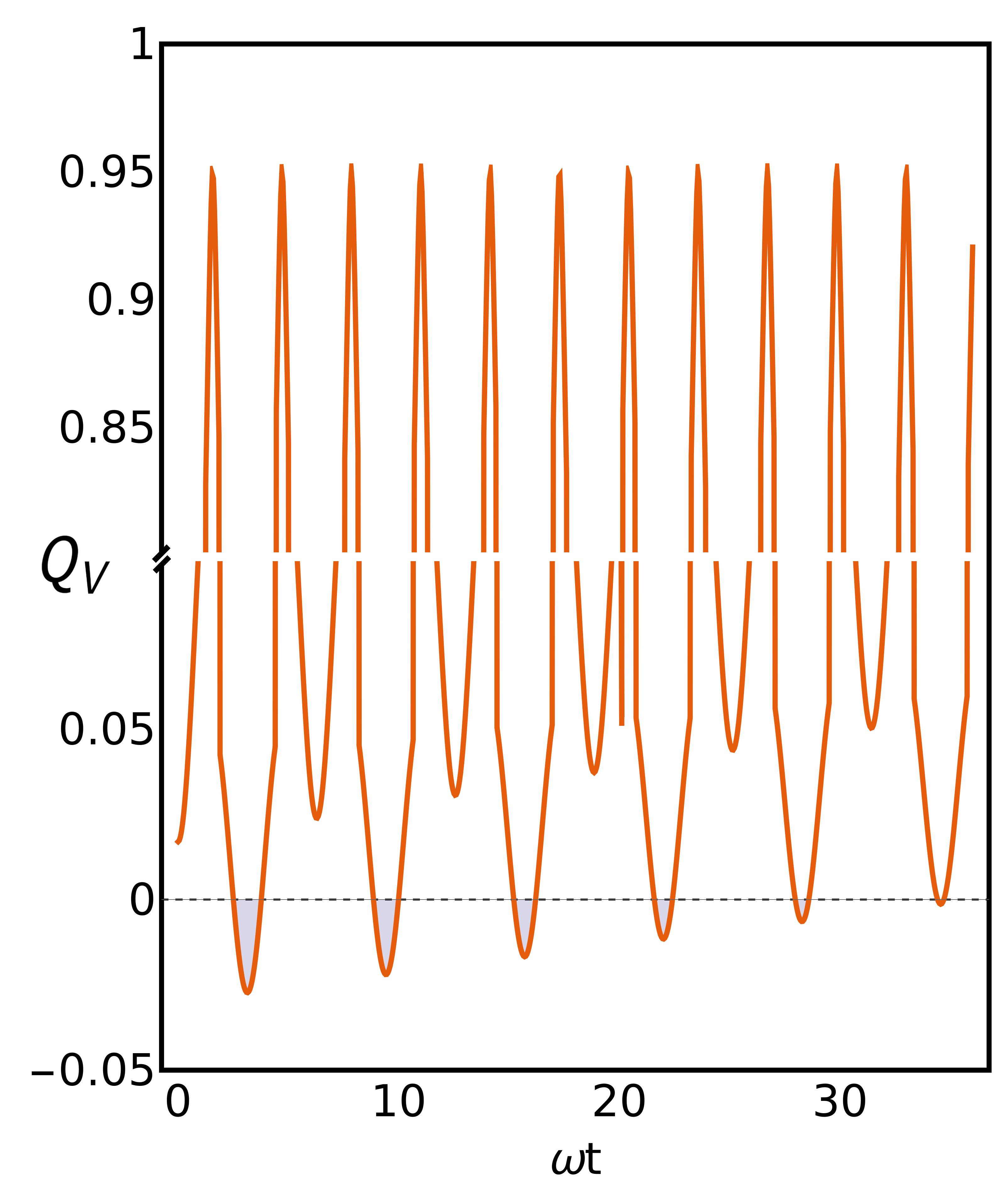}
	\caption{Non-local correlations in an OQS described by the Hamiltonian of Eq.(\ref{altmodel}), with $N=20$ and $\omega=1$, when a jump operator $\mathcal{J} = S_{y}/\sqrt{N}$ is considered. Light-blue shaded areas correspond to the negative values of the quantum violation ($Q_v$) as detected by Bell operator (\ref{bell2}). The two slanting lines on the y-axis indicate that part of the range has been omitted for the ease of readability.}
	\label{fig:dy2}
\end{figure}

We study the robustness of non-locality in an OQS also in an out of equilibrium scenario, by inspecting the time evolution of the system when dissipation is taken into account.
Referring to the Hamiltonian of Eq.(\ref{lmg2}), we consider the simpler case $h=0$, i.e.,
\begin{equation}
\label{altmodel}
H_{S} = -\omega S_{z}^{2} + \frac{\omega N}{2}\left(\frac{N}{2}+1\right)~,
\end{equation} 
\noindent with $\omega=J/N$.
Starting from a local state of the form $\rho_{S}(0)= \ketbra{D^{N}_{N/2}}{D^{N}_{N/2}}+\ketbra{D^{N}_{N/2+1}}{D^{N}_{N/2+1}}+\ketbra{D^{N}_{N/2}}{D^{N}_{N/2+1}}+\mbox{h.c.}$, we consider the evolution of the system in the presence of dissipation. In Fig.\ref{fig:dy2} we plot the quantum violation detected by means of Eq.(\ref{bell2}) when a jump operator $\mathcal{J}=S_{y}/\sqrt{N}$ in Eq.(\ref{gksl}) is considered, with $\gamma=0.001$. Despite the presence of the dissipation, non-local correlations arise periodically in the system and survive for a certain time. Moreover, the spin squeezing criterion and the analysis of the concurrence attest the presence of entanglement in the dynamical state at all times. 

Finally, we explore the scenario in which the many-body system \text{S} undergoes frequent measurements. From a physical point of view, this setting can be interpreted as the attempt of an eavesdropper to gain information on S while remaining undetected. In this case, the equation for the evolved state $\rho_\text{S}(t)$ is given by Eq.(\ref{measurement}). Our analysis is structured as follows: first, we consider an initial state  $\rho_\text{S}(0)$ derived by numerically solving Eq.(\ref{gksl}); second, we compute its dynamical evolution under repeated measurements with an operator $\mathcal{M}=S_z$ by means of Eq.(\ref{measurement}); third, we plot the non-local correlations detected by the Bell operator of Eq.(\ref{bell2}) for different values of the rate $\kappa$ as a function of the probability $p = \kappa t$. 
\begin{figure}[ht]
	\includegraphics[width=8.6cm,height=10.05cm]{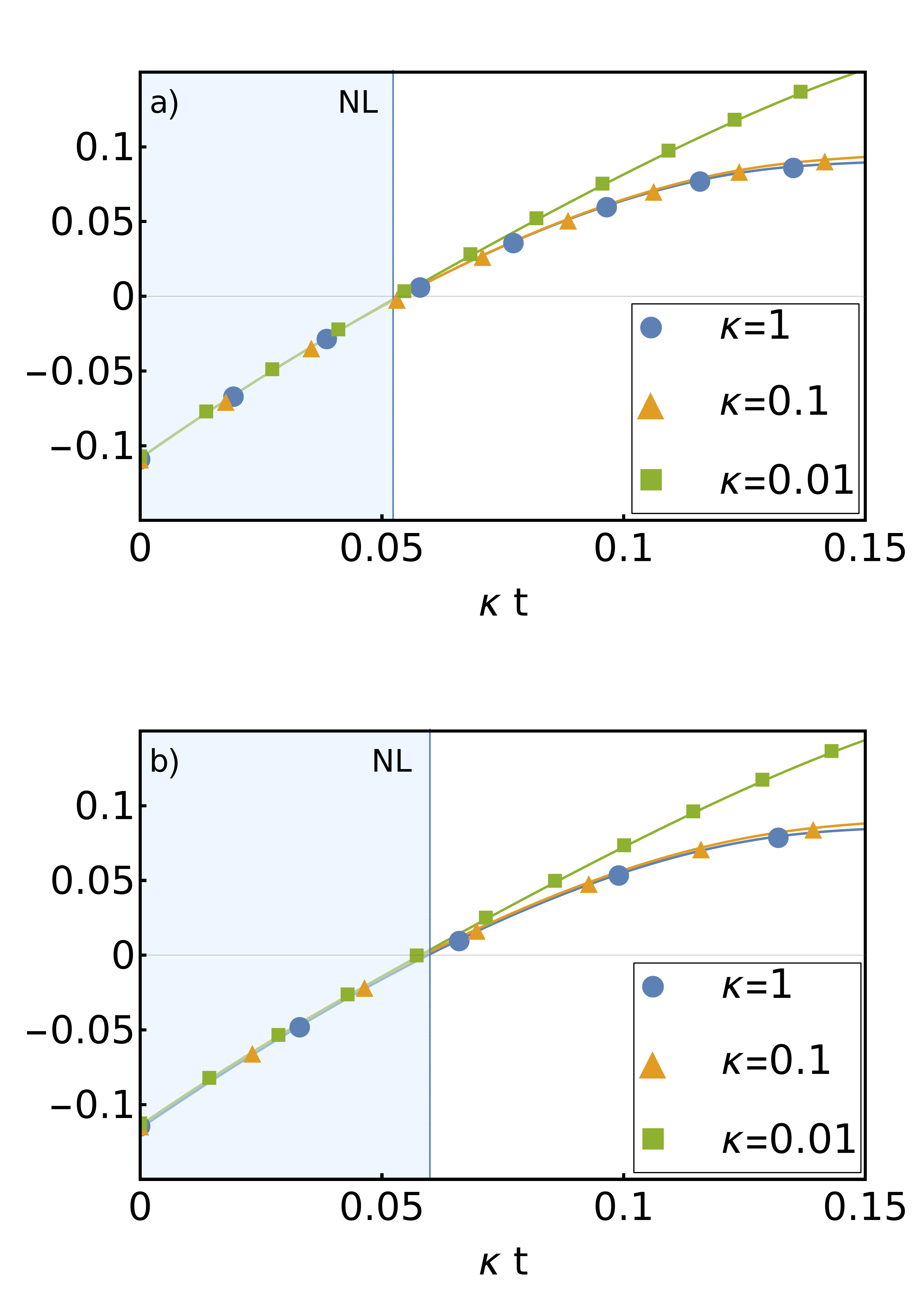}
	\caption{Non-local correlations in a many-body system undergoing a repeated measurement $\mathcal{M}=S_{z}$ at rate $k$. Blue dots correspond to $\kappa =1$; orange triangles to $\kappa = 0.1$ and green squares to $\kappa = 0.01$.  The initial state has been chosen as a: a) thermal; b) non-thermal state. NL denotes the non-local region detected by the Bell operator of Eq.(\ref{bell2}).}
	\label{fig:me}
\end{figure}
Our results are shown in Fig.\ref{fig:me} where we consider two different initial states: the thermal steady-state for $\beta=30$ (Fig.\ref{fig:me}a); and the non-thermal steady-state for $\zeta=0.01$ (Fig.\ref{fig:me}b). In both cases we choose $N=30$, $h=0.02$ and $J=1$.
Our analysis shows that, except for a slight difference in the survival time, the trend is basically the same in the two scenarios. Moreover, we observe that, despite the different values of the measurement rate $\kappa$, the slope is identical for all the curves in the non-local region. This can be understood as follows: for sufficiently short times such that $J t, h t, \kappa t \ll 1$, one can expand the map generated by Eq.(\ref{measurement}) as
\begin{align}
\label{eq:approx}
\partial_t \rho_\text{S}(t) &= \rho_\text{S}(0) -i[H_\text{S},\rho_\text{S}(0)] + \kappa t \left(\sum_k \Pi_k \rho_\text{S}(0) \Pi_k^\dagger - \rho_\text{S}(0) \right) \nonumber\\
&= (1-\kappa t) \rho_\text{S}(0) + \kappa t \sum_{k} \Pi_k \rho_\text{S}(0) \Pi_k^\dagger~,
\end{align}
where we have assumed that $[H_\text{S},\rho_\text{S}(0)]=0$.\\
Hence, with the aid of Eq.(\ref{eq:approx}), the state $\rho_{S}(t)$ can be cast as a convex combination, with probability $p=\kappa t$, of the initial state $\rho_\text{S}(0)$ and the dephased state $\bar{\rho}_\text{S}(0) = \sum_k \Pi_k \rho_\text{S}(0) \Pi_k^\dagger$, i.e., $\rho_\text{S}(t) \approx (1-p) \rho_\text{S}(0) + p \bar{\rho}_\text{S}(0)$. Since we have neglected higher order terms in the exponential map, this result is valid only for $\kappa t\ll 1$, a condition which also guarantees the probability $p$ to be bounded.

\textit{Conclusions}.---We have demonstrated not only that non-local correlations can survive in many-body systems in the presence of noise, but also that they can be robust against it. Non-locality in a many-body system coupled to a thermal bath has already been studied in \cite{fadel2018bell}, although their analysis was restricted to the stationary regime. On the contrary, in our work we have considered a dynamical scenario showing that, starting from a local state $\rho_{S}(0)$, non-local correlations arise in the state $\rho_{S}(t)$ despite the presence of either thermal or non-thermal dissipation. Moreover, we have compared the presence of non-locality with the emergence of entanglement by means of either the concurrence and the spin squeezing criterion. In both cases entanglement appears in the OQS significantly earlier than the violation of the Bell inequality, a feature that provides further evidence of the inequivalence between entanglement and non-locality. Finally, we have investigated a scenario in which a many-body system is repeatedly measured. This setup is particularly interesting when interpreted as the attempt of an eavesdropper to gain information on the system. Interestingly, our results show that the effect of the frequent measurements is to disrupt non-locality in the OQS for small time scales while also demonstrating a finite range where non-locality survives, a fact which should be reconsidered in the assessment of the security of the quantum cryptography protocols in noisy scenarios. \\
\textit{Acknowledgements}.---We acknowledge financial support from Spanish Agencia Estatal de Investigación PID2019-107609GB-100, Generalitat de Catalunya CIRIT 2017-SGR-1127, QuantumCAT 001-P-001644, co-funded by the European Union Regional Development Fund within the ERDF Operational Program of Catalunya 2014-2020. A.A. acknowledges financial support by the Austrian Science Fund (FWF) project P 33730-N.
%We acknowledge financial support from ERC-AdG NOQIA; Spanish Agencia Estatal de Investigación (AEI): Severo Ochoa Grant No. CEX2019-000910-S, PID2019-107609GB-100, PID2019-106901GB-I00/10.13039 / 501100011033, MINCIN-EU QuantERA MAQS PCI2019-111828-2, 10.13039/501100011033; Generalitat de Catalunya: AGAUR 2017-SGR-1341,2017-SGR-1127, CERCA Program and  QuantumCAT 001-P-001644, QuantumCAT U16-011424 co-funded by ERDF Operational Program of Catalonia 2014-2020; Fundaci\'o Privada Cellex; Fundaci\'{o} Mir-Puig; EU Horizon 2020 FET-OPEN OPTOLogic: Grant No. 899794; and the National Science Centre Poland-Symfonia Grant No. 2016-20-W-ST4-00314.
\bibliography{non_local_oqs}
\bibliographystyle{ieeetr}
\end{document}